\documentclass[preprint,12pt,doublespace]{elsarticle}

\usepackage[english]{babel}
\usepackage{nomencl} 

\usepackage[a4paper,top=2.54cm,bottom=2.54cm,left=2.54cm,right=2.54cm,marginparwidth=1.75cm]{geometry}

\usepackage{amsmath}
\usepackage{amssymb}
\usepackage{bm}
\usepackage{graphicx}
\usepackage[colorinlistoftodos]{todonotes}
\usepackage[colorlinks=true, allcolors=black]{hyperref}
\usepackage[superscript,biblabel]{cite}

\usepackage[rightcaption]{sidecap}

\usepackage{natbib}

\usepackage{acronym}

\usepackage{multirow}

\usepackage{caption}
\usepackage{subcaption}

\usepackage{ragged2e}

\journal{arXiv}

\begin{document}

\title{Calibration and surrogate model-based sensitivity analysis of crystal plasticity finite element models}

\author[address]{Hugh Dorward\corref{mycorrespondingauthor}}
\cortext[mycorrespondingauthor]{Corresponding author}
\ead{hugh.dorward@bristol.ac.uk}
\author[address]{David M. Knowles}
\author[address1]{Eralp Demir}
\author[address]{Mahmoud Mostafavi}
\author[address]{Matthew J. Peel}

\address[address]{Department of Mechanical Engineering, University of Bristol, Bristol, BS8 1TR, UK.}
\address[address1]{Department of Engineering Science, University of Oxford, Parks Road, Oxford OX1 3PJ, UK.}

\begin{abstract}
Crystal plasticity models are a powerful tool for predicting the deformation behaviour of polycrystalline materials accounting for the underlying grain morphology and texture. These models typically have a large number of parameters, an understanding of which is required to effectively calibrate and apply the model. This study presents a structured framework for the global sensitivity analysis, using Sobol indices, of the effect of crystal plasticity parameters on model outputs. Due to the computational cost of evaluating the crystal plasticity model multiple times within a finite element framework, a Gaussian process regression surrogate model was constructed and used to conduct the sensitivity analysis. Influential parameters from the sensitivity analysis were carried forward for calibration using both a local Nelder-Mead and  global differential evolution optimisation algorithm. The results show that the surrogate based global sensitivity analysis is able to efficiently identify influential crystal plasticity parameters and parameter combinations. With a reduced parameter set, both the Nelder-Mead algorithm with multiple runs and the differential evolution algorithm were able to find an optimal parameter set which gives a close calibration of a simulated tensile curve to experimental data. However, only the differential evolutionary algorithm was able to reliably find the global optimum due to the presence of local minima in the calibration objective function.

\section*{Keywords}
\noindent crystal plasticity, sensitivity analysis, Gaussian process, surrogate model, calibration

\end{abstract}

\maketitle

\newpage

\section{Introduction} \label{sect:Introduction}
Steels are one of the most common engineering materials and have been extensively investigated. Nevertheless, there are safety-critical applications, such as the nuclear and aerospace industries\cite{monnet2019prediction,yaghoobi2021crystal}, that necessitate a detailed understanding of deformation and failure and how these relate to the material microstructure. Meso-scale crystal plasticity modelling is a well-established technique that has gathered growing interest for application within industrial contexts. This interest is partly due to increases in computational performance, which makes evaluation of larger models more feasible. Application of crystal plasticity models within both finite element and spectral frameworks allow simulations of representative volume elements of a polycrystalline material, which account for features such as grain size, morphology and texture \cite{roters2010overview}. 

Different crystal plasticity models can be broadly split into those adopting phenomenological constitutive equations, \cite{agius2020microstructure} and those adopting physically based constitutive equations \cite{arsenlis1999crystallographic,bayley2006comparison}. Both approaches typically have a number of parameters, though generally fewer in the case of phenomenological formulations. These need to be determined either directly from experiment or by fitting the models to experimental data for specific materials. Some models can be quite complex, incorporating a number of evolutionary equations for different state variables. Therefore, it is important to understand the effect of specific crystal plasticity parameters on the model behaviour and to ensure careful and consistent calibration of these parameters.

Undertaking sensitivity analyses with respect to crystal plasticity constitutive law parameters offers the advantage of highlighting the key parameters that are most influential on the model response \cite{agius2020microstructure}. Previous studies have quantified the sensitivity of parameters of a multi-component Armstrong-Frederick model to the accuracy of prediction of the cyclic hysteresis response \cite{agius2017sensitivity,agius2018optimising}. Other studies have investigated the effect of crystal plasticity parameters on stress-strain response in a more qualitative manner by adjusting single parameters at a time and looking at the effect this has on the stress strain response for each parameter \cite{taylor2020investigation,kulkarni2023sensitivity}. Few attempts, however, have been made to conduct comprehensive sensitivity analyses on crystal plasticity models, particularly analyses accounting for higher order interactions between crystal plasticity parameters.

Many approaches exist for sensitivity analysis of models and systems. One-at-a-time sensitivity analyses, such as the Morris method, \cite{morris1991factorial} involve changing each parameter individually while keeping other parameters constant. These can be completed with a relatively small number of complex model evaluations \cite{chakraborty2017evaluation}. More comprehensive global methods, however, require a larger number of model evaluations \cite{iooss2015review} but are capable of analysing the higher order combined interactions between parameters. The large number of simulations required for global methods such as Sobol sensitivity analysis \cite{sobol2001global} poses a barrier to application to complex, computationally expensive crystal plasticity models. Typically, tens of thousands of simulations are required \cite{iooss2019advanced}. A solution to this problem has been demonstrated in studies that use surrogate models to emulate the complex model response and then perform the sensitivity analyses on these surrogate models. For example, computation of Sobol sensitivity indices has been demonstrated through the use of Gaussian process regression models \cite{iooss2019advanced} and polynomial chaos expansion methods \cite{crestaux2009polynomial}. These studies demonstrate a framework for applying global sensitivity analysis to complex, computationally expensive models with a large number of input parameters.

The computational expense of crystal plasticity models also makes calibration of model parameters challenging. A number of studies still rely on manual calibration and trial and error to find a calibrated set of parameters \cite{verma2016crystal}. Increasingly, optimisation algorithms have been used to find optimal crystal plasticity model parameters. These can be divided into local optimisation algorithms, such as the Nelder-Mead algorithm, \cite{chakraborty2017evaluation,engels2019parameterization} and global algorithms, such as evolutionary algorithms \cite{prithivirajan2018role,kapoor2021modeling}. While local methods can be relatively efficient, there is no guarantee they will find a global solution to the calibration problem as they can get stuck in local minima. Therefore, multiple restarts are often required from different starting locations. Conversely, global methods are designed to find the global solution of the problem, usually through the introduction of stochasticity to the search criteria, but they often require a larger number of model evaluations. The efficiency of these optimisation algorithms, however, may be highly dependent on the optimisation hyperparameters selected by the user. Here, hyperparameters refers to parameters which control the search strategy of the optimisation algorithm. This terminology has been adopted in this paper to distinguish these hyperparameters from crystal plasticity parameters. Few studies in the literature report the hyperparameters used or compare these to the optimisation performance for different sets of hyperparameters. Similarly, justification is rarely given in the choice between different local and global optimisation algorithms.

In this study, a framework is developed for the sensitivity analysis and calibration of a strain gradient crystal plasticity model allowing selection and calibration of the most influential parameters on model behaviour. Firstly, a surrogate model based sensitivity analysis is undertaken to determine the Sobol sensitivity indices for the crystal plasticity parameters with respect to various macroscopic quantities of interest. This sensitivity study is used to inform a reduced set of crystal plasticity parameters for calibration. A local Nelder-Mead calibration procedure is conducted to determine the optimal parameter set. This is compared with a global differential evolution procedure regarding computational efficiency and ease of use.



\section{Strain gradient crystal plasticity model and parameters} \label{sect:Crystal_plasticity}
The crystal plasticity model used in this study is a dislocation-based strain gradient model. The model was implemented in ABAQUS finite element software \cite{abaqus} using the deformation gradient formulation and implicit time integration procedure of Kalidindi et al. \cite{kalidindi1992crystallographic}. The flow laws, hardening laws, and evolutionary equations specific to this strain gradient model are provided in Section \ref{sect:strain_grad} below.

\subsection{Crystal plasticity constitutive equations} \label{sect:strain_grad}

The sliprate on each slip system allows calculation of the plastic deformation gradient according to the procedure in Kalidindi et al. \cite{kalidindi1992crystallographic}. This is a function of the resolved shear stress, $\tau^{\alpha}$, and is calculated per slip system, $\alpha$, according to the power law relationship in Equation \ref{eq:Flow_rule} following \cite{hutchinson1976bounds,peirce1982analysis,peirce1983material}:

\begin{equation} \label{eq:Flow_rule}
\dot{\gamma}^\alpha   \;=\; \dot{\gamma}_0 \, \left|  \dfrac{ \tau^\alpha  }{\tau_c^\alpha} \right|^{1/m} \! \text{sign}(\tau^\alpha).
\end{equation}

\noindent The constants $\dot{\gamma}_0$ and $m$ in this equation represent the initial slip rate and the rate insensitivity exponent respectively. The hardening law calculates the critical resolved shear stress, $\tau_c^{\alpha}$, and is dependent on the evolution of statistically stored dislocation (SSD) density and geometrically necessary dislocation (GND) density. GNDs are required to maintain compatibility due to deformation in a crystal lattice and SSDs are dislocations which become stored by random interactions within the crystal lattice \cite{fleck1994strain}. The critical shear stress on each slip system is split into two terms. One is given by the initial threshold shear stress, $\tau_c^0$, and the other (i.e. work hardening) by the contribution of the SSD density, $\varrho_{SSD}^\alpha$, and the screw and edge GND densities, $\varrho_{GND,s}^\alpha$ and $\varrho_{GND,e}^\alpha$ \cite{ashby1970deformation}. In Equation \ref{eq:Hardening_rule}, $G$ is the shear modulus, $b$ is the Burgers vector magnitude and $C$ is a proportionality constant.

\begin{equation} \label{eq:Hardening_rule}
\tau_c^\alpha \;=\; \tau_{c}^0 \;+\; C \, G \, b \sqrt{\sum_{\beta=1}^{N_{\mathrm{slip}}}\chi^{\alpha\beta} \, \left(\varrho_{SSD}^\beta +  \left| \varrho_{GND,s}^\beta \right| +  \left| \varrho_{GND,e}^\beta \right|\right)}.
\end{equation}

The interaction matrix, $\chi^{\alpha\beta}$, determines cross hardening interactions between planes. In this study, a hardening factor of unity for co-planar slip systems is assumed, with a constant value, $q$, otherwise:

\begin{equation}
    \chi^{\alpha\beta} \,=\, \begin{bmatrix}
\mathbf{1} & q \mathbf{1} & q \mathbf{1} & q \mathbf{1}\\
q\mathbf{1} & \mathbf{1} & q \mathbf{1} & q \mathbf{1}\\
q\mathbf{1} & q \mathbf{1} & \mathbf{1} & q \mathbf{1}\\
q\mathbf{1} & q \mathbf{1} & q \mathbf{1} & \mathbf{1}\\
\end{bmatrix} \text{, where; }
\mathbf{1} \,=\, \begin{bmatrix}
1 & 1 & 1\\
1 & 1 & 1\\
1 & 1 & 1\\
\end{bmatrix}.
\end{equation}

As the total dislocation density increases, the number of dislocations trapped as SSDs increases. Therefore, the SSD density increases as a function of the total dislocation density \cite{arsenlis2002modeling}. This is balanced with a negative term in proportion with the SSD density to account for the athermal interaction and annihilation of SSDs with the critical annihilation distance represented by $y_c$ according to Equation \ref{eq:SSD_evolution}. 

\begin{equation} \label{eq:SSD_evolution}
\dot{\varrho}_{SSD}^\alpha \;=\; \left(K \,\sqrt{\varrho_{SSD}^\alpha + \left| \varrho_{GND,e}^\alpha \right|+ \left| \varrho_{GND,s}^\alpha \right| } \,-\, 2 y_c \, \varrho_{SSD}^\alpha\right) \, \dfrac{\left|\dot{\gamma}^\alpha\right|}{b}.
\end{equation}

GNDs are required to maintain compatibility in the crystal lattice under deformation and are therefore related to the local gradient of strain on a given slip system according to Equations \ref{eq:GNDe_evo} and \ref{eq:GNDs_evo} depending on whether they are in edge or screw configuration. $\mathbf{s}^\alpha$ is the slip direction for edge dislocations which is perpendicular to the dislocation line vector, whereas $\mathbf{t}^\alpha$ represents the direction for screw dislocations which is parallel to the dislocation line vector.

\begin{equation} \label{eq:GNDe_evo}
\dot{\varrho}_{GND,e}^\alpha \;=\;  - \dfrac{1}{b}\, \nabla \dot{\gamma}^\alpha \cdot \mathbf{s}^\alpha.
\end{equation}

\begin{equation} \label{eq:GNDs_evo}
\dot{\varrho}_{GND,s}^\alpha \;=\; \dfrac{1}{b}\, \nabla \dot{\gamma}^\alpha \cdot \mathbf{t}^\alpha.
\end{equation}

\section{Methods} \label{sect:Methods}

The purpose of the sensitivity analysis methodology presented in Section \ref{sec:method_SA} is to determine the effect of both individual crystal plasticity parameters and combined interactions of multiple parameters on various macroscopic quantities of interest. This sensitivity is used to inform the parameters selected for calibration in the local and global optimisation methods described in Section \ref{sec:methods_calib}.

The aim of the optimisation algorithms is to minimise the difference between the experimental data and the simulated curve. The normalised mean squared error (NMSE) shown in Equation \ref{eq:objective_function} is used to quantify the difference between the simulated tensile curve for a given combination of parameters, $\theta$, compared to the experimental results.

\begin{equation} \label{eq:objective_function}
    f(\mathbf{\theta}) = \frac{1}{N}\sum_{i=0}^N \left(\sigma_i(\mathbf{\theta})^{(\mathrm{sim})} -  \sigma_i^{(\mathrm{exp})}\right)^2
\end{equation}

\begin{figure}[h!]
    \centering
    \includegraphics[width=0.55\textwidth]{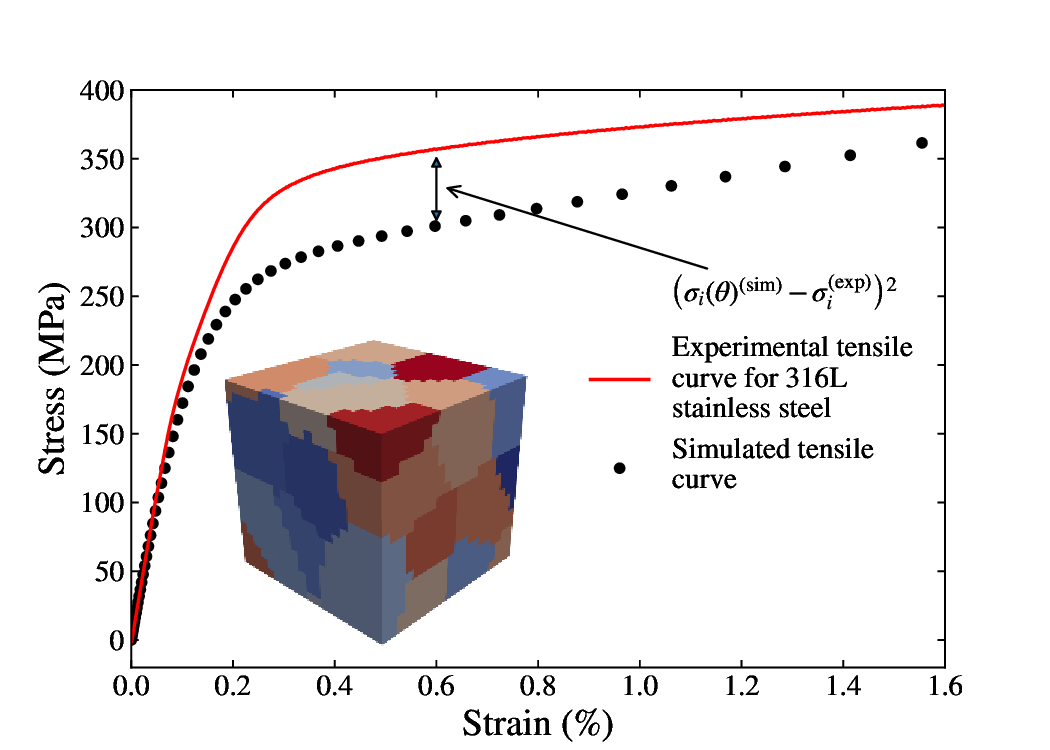}
    \caption{Experimental tensile curve for 316L stainless steel showing the calculation of mean-squared error for each datapoint from a crystal plasticity simulation with parameters $\theta$. The polycrystal SVE used for the crystal plasticity finite element simulation is shown on the figure.}
    \label{fig:ExpCurve}
\end{figure}

\noindent This equation is demonstrated visually in Figure \ref{fig:ExpCurve}, with an example simulated curve and the experimental data used. The calibration problem can therefore be formulated as a minimisation problem as in Equation \ref{eq:Calibration}, where $d$ is the number of parameters which are free to vary for the calibration and $f(\theta)$ is the objective function.

\begin{equation} \label{eq:Calibration}
    \min f(\mathbf{\theta}) \quad \mathrm{where} \quad \mathbf{\theta} \in \mathbb{R}^d
\end{equation}

The experimental data used for this calibration was from the tensile test of a 316L stainless steel specimen at room temperature by Mokhtarishirazabad et al. \cite{mokhtarishirazabad2021evaluation,mokhtarishirazabad2023predicting} according to ASTM E8. The tensile specimen was taken from a 25mm thick plate. In the current study, DREAM.3D software\cite{groeber2014dream} was used to generate the statistical volume element (SVE) used for the crystal plasticity finite element simulations. Initially, ten SVEs were generated with an untextured, equiaxed grain structure with an average grain size of $20\pm8\mu$m. These statistics were used to replicate the microstructure observed in the experimental specimen. To account for statistical differences in the response of different SVEs, the SVE with the response closest to the mean was used for further analysis and is shown in Figure \ref{fig:ExpCurve}.

\subsection{Sensitivity analysis - Sobol sensitivity indices}
\label{sec:method_SA}

Sobol sensitivity indices are the outputs of a global sensitivity analysis method which aims to quantify the proportion of variance explained by any combination of parameters. Given a function or model, $y=f(\mathbf{x})$, where $\mathbf{x}$ is a vector of input parameters and $\mathbf{x} \in \mathbb{R}^d$, a functional decomposition can be given as in Equation \ref{eq:func_decomp}:

\begin{equation} \label{eq:func_decomp}
    f(\mathbf{x}) = f_0 + \sum^d_{i=1}f_i(x_i) + \sum^d_{i<j}f_{ij}(x_i,x_j) + ... + f_{12...d}(x_1,x_2,...,x_d).
\end{equation}

\noindent Similarly, when the input parameters are independent, the variance of the function $Y$ can be decomposed into the sum of effects due to different parameters as in Equation \ref{eq:var_decomp}:

\begin{equation} \label{eq:var_decomp}
    \mathrm{Var}(Y) = \sum^d_{i=1}D_i(Y) + \sum^d_{i<j}D_{ij}(Y)+...+D_{12...d}(Y).    
\end{equation}

\noindent Here, $D$ represents the variance due to a specific combination of parameters as shown for the first and second order terms in Equations \ref{eq:Di} and \ref{eq:Dij}:

\begin{equation} \label{eq:Di}
    D_i(Y) = \mathrm{Var}[\mathbb{E}(Y|x_i)],
\end{equation}

\begin{equation} \label{eq:Dij}
    D_{ij}(Y) = \mathrm{Var}[\mathbb{E}(Y|x_i,x_j)] - D_i(Y) - D_j(Y).
\end{equation}

\noindent Sobol indices can then be calculated for each combination of parameters by finding the proportion of variance for each combination of parameters to the total variance as in Equations \ref{eq:Si} and \ref{eq:Sij} for the first and second order indices:

\begin{equation}\label{eq:Si}
    S_i = \frac{D_i(Y)}{\mathrm{Var}(Y)}
\end{equation}

\begin{equation}\label{eq:Sij}
    S_{ij} = \frac{D_{ij}(Y)}{\mathrm{Var}(Y)}
\end{equation}

In practice, visualising all the higher order indices is challenging due to the way the number of indices quickly increase as the dimension of input variables increases. Instead, it is common to specify the total Sobol index, $S_{T,i}$, which is the sum of all of the indices which contain an effect from the $i$\textsuperscript{th} parameter. The computational design of experiment to give the set of parameter values required for the analysis of Sobol indices was determined by the method of Campolongo et al. \cite{campolongo2011screening}. The SALib library \cite{Herman2017,Iwanaga2022} was used within Python for this study for the design of experiment and computation of Sobol indices.

\subsection{Gaussian process surrogate model for sensitivity analysis}
As discussed previously, the number of model evaluations required for the global sensitivity analysis is very large meaning a surrogate model is required to emulate the response of the crystal plasticity model. Surrogate models are typically machine learning regression algorithms which aim to learn the response of a complex function or model. Once trained, the goal of the surrogate model is to predict the output of the complex model given a new set of input points. 

In this study, Gaussian process (GP) regression is used to predict the output quantities of interest based on the crystal plasticity parameters. GP modelling is a non-parametric machine learning method where the output for a test data point located at $\mathbf{x}_*$ within the input variable space is predicted based on an observed set of training data points, $X=[\mathbf{x}_1,\mathbf{x}_2,...,\mathbf{x}_n]^{\mathrm{T}}$. The Gaussian process model is defined as a multivariate Gaussian distribution, as shown in Equation \ref{eq:GP_def}. This is defined by a mean vector, which is zero in this instance indicating that no prior knowledge is known about the trend of the function, and the covariance matrix, $K(X,X)$. Each dimension of the distribution is a random variable that represents a point in the input variable space. 
\begin{equation} \label{eq:GP_def}
Y \sim \mathcal{N}\left(
    \mathbf{0} , K(X,X) \right)
\end{equation}

Further detail in the construction of GP models for regression is provided by Williams \& Rasmussen \cite{williams2006gaussian}, however a crucial part of the model is the covariance function, $k(\mathbf{x}_i,\mathbf{x}_j)$, which governs the entries in the covariance matrix. The covariance function or kernel function describes the relationship between two data points $\mathbf{x}_i$ and $\mathbf{x}_j$, either of which can be training data where the outputs are known, or testing data where the outputs are to be predicted. The relationship between the covariance matrix and covariance function is shown in Equation \ref{eq:covar}.
\begin{equation} \label{eq:covar}
    K(X,X) = 
    \begin{bmatrix}
    k(\mathbf{x}_1,\mathbf{x}_1) & \dots &
    k(\mathbf{x}_1,\mathbf{x}_n) \\
    \vdots & \ddots & \vdots\\
    k(\mathbf{x}_n,\mathbf{x}_1) & \dots &
    k(\mathbf{x}_n,\mathbf{x}_n)    
    \end{bmatrix}
\end{equation}
Typically, this is a function of the Euclidean distance between $\mathbf{x}_i$ and $\mathbf{x}_j$, such as the radial basis function shown in Equation \ref{eq:RBF}: 

\begin{equation} \label{eq:RBF}
    k(\mathbf x_i,\mathbf x_j) = \sigma^2_f\, \mathrm{exp}(-\frac{1}{2\ell}|\mathbf x_i - \mathbf x_j|^2).
\end{equation}

\noindent The radial basis function is infinitely differentiable and is therefore suited for fitting smooth functions. As the response function for the output quantities of interest was expected to be reasonably smooth, the radial basis function was adopted for this study. To make predictions from the GP model, the conditional probability at a test point with unknown output is found. Conditional probabilities from a multivariate Gaussian distribution are straightforward to calculate \cite{williams2006gaussian}, and also follow a Gaussian distribution. This means the prediction from a GP model is a Gaussian random variable, the variance of which gives a value for the epistemic surrogate-model uncertainty in the prediction. In Equation \ref{eq:RBF}, $\sigma_f^2$ and $\ell$ are hyperparameters of the GP model which determine the behaviour of the covariance function. Point estimates for these hyperparameters can be determined using optimisation to find the \textit{maximum a posteriori} value for each hyperparameter. In this study, the PYMC library was used to construct the GP model, train the model and tune the hyperparameters \cite{salvatier2016probabilistic}. To train the model two data sets were used, the experimental design for each being determined by a space-filling Sobol sequence. The first data set consists of $20d$ (160) data points and the second is the same as the first but with an additional 160 data points. 



\subsection{Parameter calibration} \label{sec:methods_calib}

For the calibration of the crystal plasticity model, two optimisation algorithms are compared to minimise the objective function in Equation \ref{eq:objective_function}. These are the Nelder-Mead algorithm and the differential evolution algorithm. In this study, both algorithms were implemented in Python using the SciPy optimize toolbox \cite{virtanen2020scipy}. It is important, however, to understand the methodology of each algorithm so an overview is provided for each below. Familiarity with the methods allows the user to understand any limitations associated with them and informs on how to set appropriate optimisation hyperparameters.

The Nelder-Mead algorithm \cite{nelder1965simplex} is a local numerical method suited to minimisation problems. The procedure is initiated with the construction of $d$-dimensional tetrahedron, or simplex, with $d+1$ vertices, where $d$ are the dimensions of the parameter space. After creating the simplex and evaluating the objective function at each vertex, a new trial point is found by reflecting the simplex around the edge opposite the vertex with the largest error. Following this reflection, a new simplex is created through expansion or contraction depending on whether the error at the trial point is higher or lower than at the previous point. The algorithm then iterates through a series of reflections, expansions and contractions until the improvement with subsequent iterations is below a tolerance set by the user. For this method, the only inputs required by the user are the initial starting point for the algorithm and the stopping criteria. As this method is a local method, multiple runs are completed in this study, with a new random starting location for each run. This shows if the algorithm is becoming stuck in local minima and allows a more detailed comparison with the global optimisation method.

The global method used in this study is the differential evolution algorithm developed by Storn and Price \cite{storn1997differential}. This is an evolutionary algorithm where parameters are represented by vectors, unlike in genetic algorithms where parameters are converted into bit-strings. The algorithm has three key steps:
\begin{enumerate}
    \item Generation and evaluation of an initial population of trial parameter vectors.
    \item Mutation of combinations of these trial vectors, by linearly combining elements from each trial vector, to create a set of mutant vectors.
    \item Crossover of mutant and trial vectors, by randomly swapping elements, to generate a new candidate parameter set.
\end{enumerate} 
The initial set of trial vectors is generated randomly and has a user defined population size. A set of mutant vectors is created by combining linearly a subset of the trial vectors. This combination is controlled by a scalar hyperparameter, $F$, which determines how amplified the variation is in each mutant vector. There are a number of different strategies for combining trial vectors, however the 'best1bin' strategy is used here, which ensures the best performing trial vector is always involved in the combination step.

The crossover step is introduced to increase diversity of the new candidate vectors. It involves crossing over elements between the original trial vector and mutant vector to create the new candidate vector. This crossover is a random process and is controlled by a crossover constant which represents the probability of crossover for each element in the vector. The function value for each candidate vector is determined and those which outperform the corresponding trial vector are selected for the next generation. This process is repeated until a maximum number of iterations is reached, or the improvement between subsequent iterations is below a certain tolerance. 

\subsection{Crystal plasticity parameter values and bounds}

In the crystal plasticity model, 8 model parameters were changed to quantify their effect on the crystal plasticity model output. The sensitivity analysis can be expected to be dependent on the parameter bounds used in the analysis. Therefore, bounds were chosen to encompass values found in literature for 316L stainless steel. Table \ref{tab:params} shows the bounds selected for each parameter as well as values used for other crystal plasticity parameters determined from literature.

\begin{table}[h!]
\centering
\begin{tabular}{c p{0.25\textwidth} c p{0.40\textwidth}}
Parameter & Description & Magnitude & Remarks \\ \hline \hline

$C_{11}$ & & 204.6 & From the 2\textsuperscript{nd} order cubic  \\
$C_{12}$ & \RaggedRight Elastic parameters  & 137.7 &  anisotropic elasticity tensor   \\
$C_{44}$ &(GPa) & 126.6 &  \RaggedRight where Young's modulus and Poisson's ratio are determined from experimental data. \\ 
$b$ & \RaggedRight Burgers vector (nm) & 0.258 & Ref. \cite{pham2013cyclic}\\
$\dot{\gamma_0}$ & \RaggedRight Initial slip rate ($s^{-1}$) & [0.001-1.0] & \RaggedRight Varies logarithmically.\\
$m$ & \RaggedRight Rate insensitivity exponent & [0.01-1.0] & \RaggedRight Varies logarithmically\\
$\tau_c^0$ & \RaggedRight Threshold critical shear stress (MPa) & [0.0-100.0] & \RaggedRight \\
$\varrho_{\mathrm{SSD}}^0$ & \RaggedRight Initial SSD density (mm$^{-2}$) & [10\textsuperscript{6}-10\textsuperscript{8}] & \RaggedRight Varies logarithmically. \\
$C$ & \RaggedRight Geometric factor & [0.0-0.5] & \\
$K$ & \RaggedRight Mean free path hardening factor (mm$^{-1}$) & [0.02-0.3] & \RaggedRight  \\
$y_c$ & \RaggedRight Annihilation radius (mm) & [0.0-$10.0\times 10^{-6}$] & \\
$q$ & \RaggedRight Latent hardening constant & [1.0-1.4] & \\

\end{tabular}

\caption{\label{tab:params} List of crystal plasticity parameter values and the bounds used for the global Sobol sensitivity analysis.}
\end{table}

\section{Results and Discussion} \label{sect:Results_Discussion}
The following sections present results for the sensitivity analysis and calibration of the crystal plasticity model. Section \ref{sec:results_SA} shows the validation of the GP surrogate model used in calculating the Sobol sensitivity indices and the results of the sensitivity analysis itself. A comparison is then provided in Section \ref{sec:results_calib} between the Nelder-Mead algorithm and the differential evolution algorithm. The reduced set of parameters used in the calibration are informed by the Sobol sensitivity analysis.
\subsection{Global Sobol sensitivity analysis} \label{sec:results_SA}
As discussed in Section \ref{sec:method_SA}, a Gaussian process regression model was implemented as a surrogate for the crystal plasticity model linking model parameters in Table \ref{tab:params} to macroscopic quantities of interest. The quantities of interest used as outputs of the model in this instance were the yield stress of the tensile curve, $\sigma_y$, the difference between the maximum stress of the tensile curve (at 1.5\% strain) and the yield stress, $\sigma_{\mathrm{max}}-\sigma_y$, and the normalised mean squared error (NMSE) of the simulated tensile curve with respect to experimental data for 316L stainless steel. Each crystal plasticity model evaluation took approximately 15 minutes using 8 PC processors, however by comparison, each surrogate model evaluation took only a fraction of second. This allowed the Sobol sensitivity analysis to be conducted on the surrogate model which was orders of magnitude faster to evaluate. 

\begin{figure}[h!]
    \centering
    \includegraphics[width=1\textwidth]{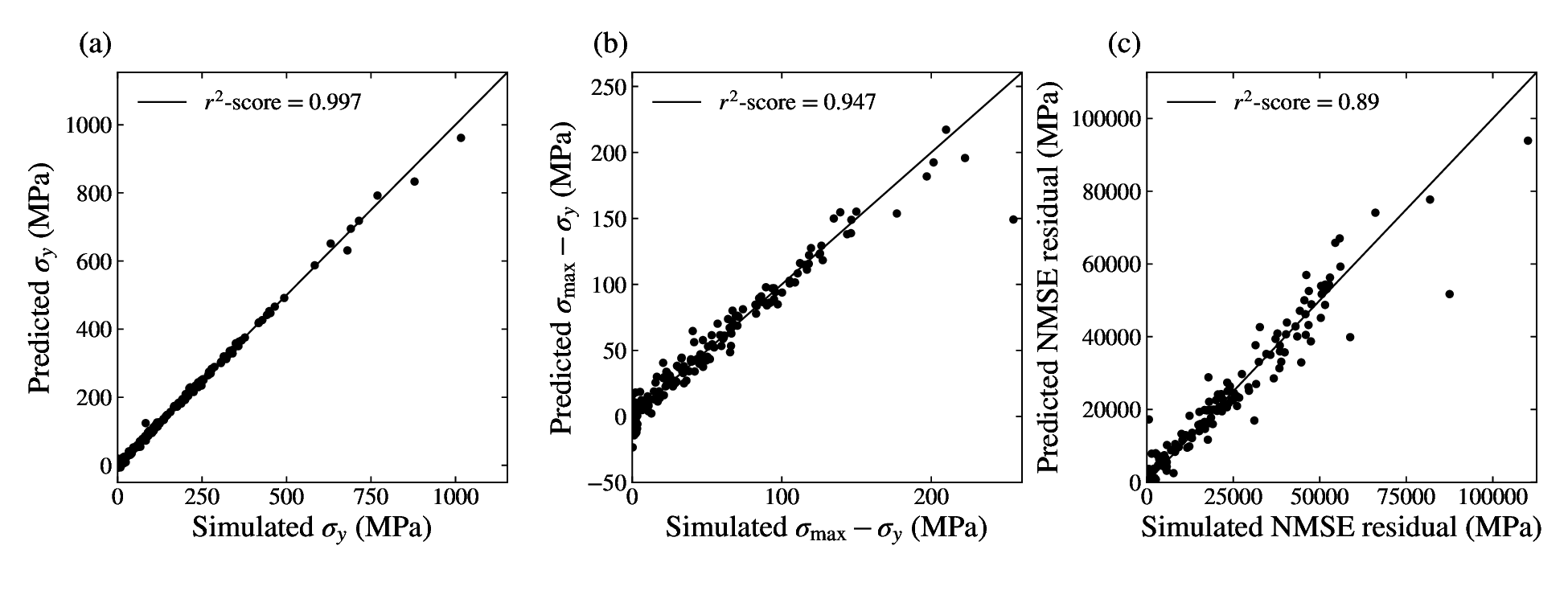}
    \caption{Results of leave-one-out cross-validation on each Gaussian process model trained on a dataset of 160 datapoints for a) the prediction of yield stress, b) the prediction of the difference between yield stress and maximum stress and c) the prediction of the NMSE residual with respect to experimental data. A straight line shows the line of unity between the simulated and predicted data.}
    \label{fig:LOOCV160}
\end{figure}

To ensure accuracy of the sensitivity analysis, the surrogate models predicting the quantities of interest must show good predictive accuracy. For this reason, leave-one-out cross-validation (LOOCV) was undertaken to validate each surrogate model. This method involves withholding one datapoint for model testing and then training a model on the remaining data. The surrogate model prediction for the withheld data point can then be compared to the actual result. This is repeated for the whole dataset, training a new model each time, so that each datapoint is used for validation once. The results of this validation for the models trained on 160 datapoints are shown in Figure \ref{fig:LOOCV160}. 

\begin{figure}[h!]
    \centering
    \includegraphics[width=1\textwidth]{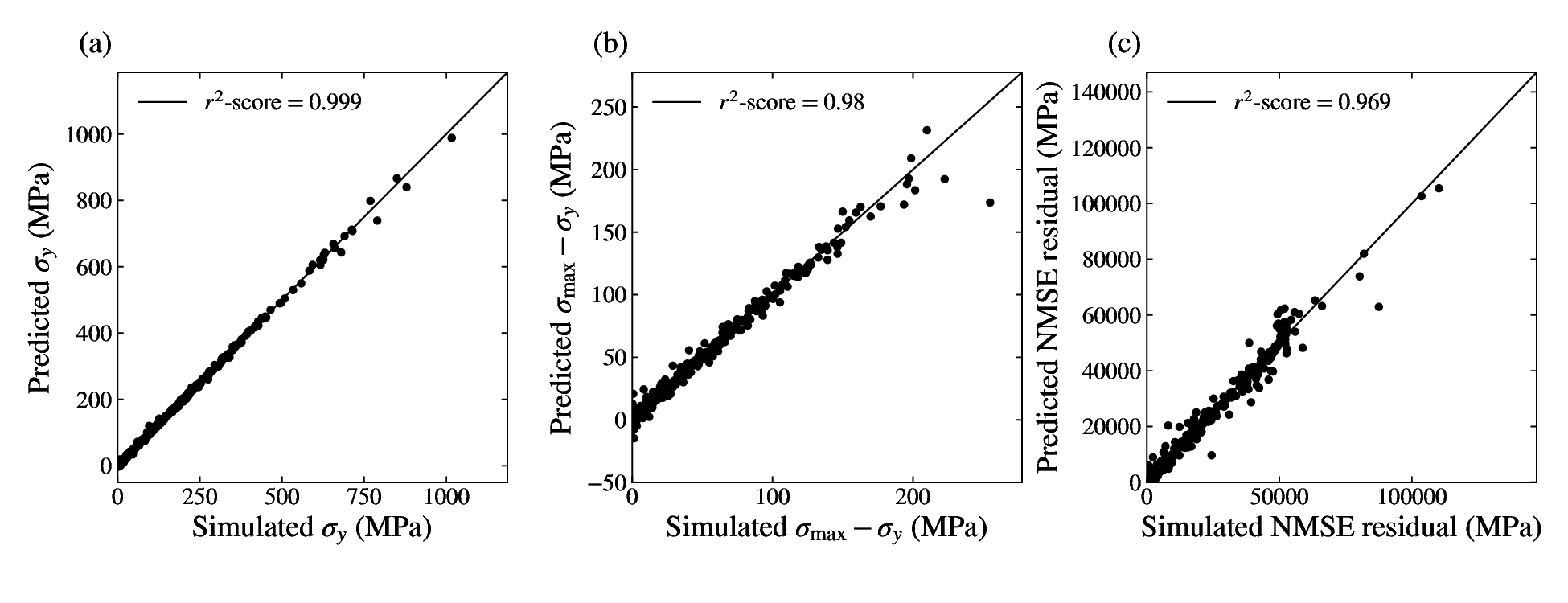}
    \caption{Results of leave-one-out cross-validation on each Gaussian process model trained on a larger dataset of 320 datapoints for a) the prediction of yield stress, b) the prediction of the difference between yield stress and maximum stress and c) the prediction of the NMSE residual with respect to experimental data. A straight line shows the line of unity between the simulated and predicted data.}
    \label{fig:LOOCV320}
\end{figure}

The predictive accuracy of the models for predicting the maximum stress and the difference between maximum stress and yield stress are very good, highlighted by the coefficient of determination of each shown in the figure. It is clear, however, that there is much higher uncertainty in the prediction of the NMSE residual. For this reason, the models were retrained on an extended dataset consisting of 320 points to improve prediction accuracy. The LOOCV results for this extended dataset are shown in Figure \ref{fig:LOOCV320} showing better predictive accuracy for all three models.

\begin{figure}[h!]
    \centering
    \includegraphics[width=1\textwidth]{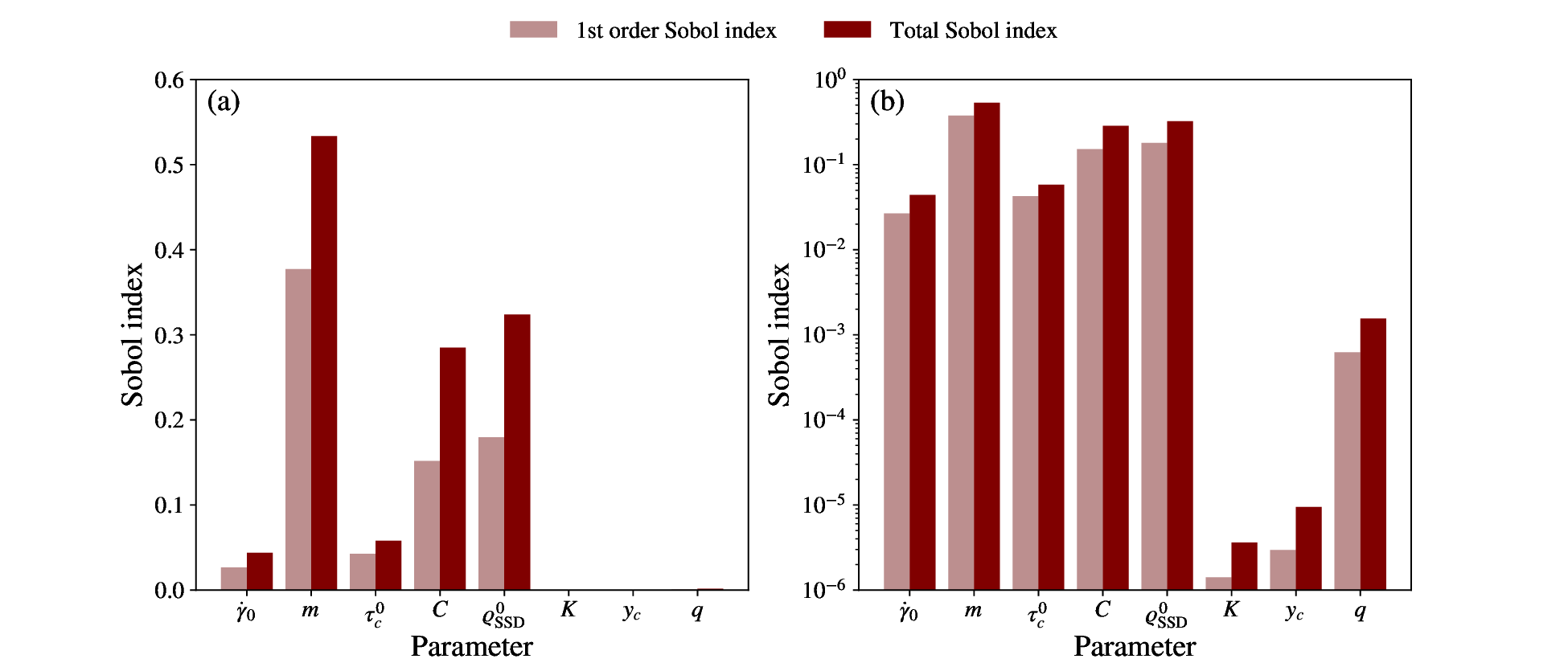}
    \caption{First order and total Sobol indices for the effect of crystal plasticity parameters on yield stress, $\sigma_y$, of the simulated tensile curve. Indices are shown on a) a linear scale and b) a logarithmic scale to show the difference between lower values.}
    \label{fig:Si_yield_stress}
\end{figure}

\begin{figure}[h!]
    \centering
    \includegraphics[width=1\textwidth]{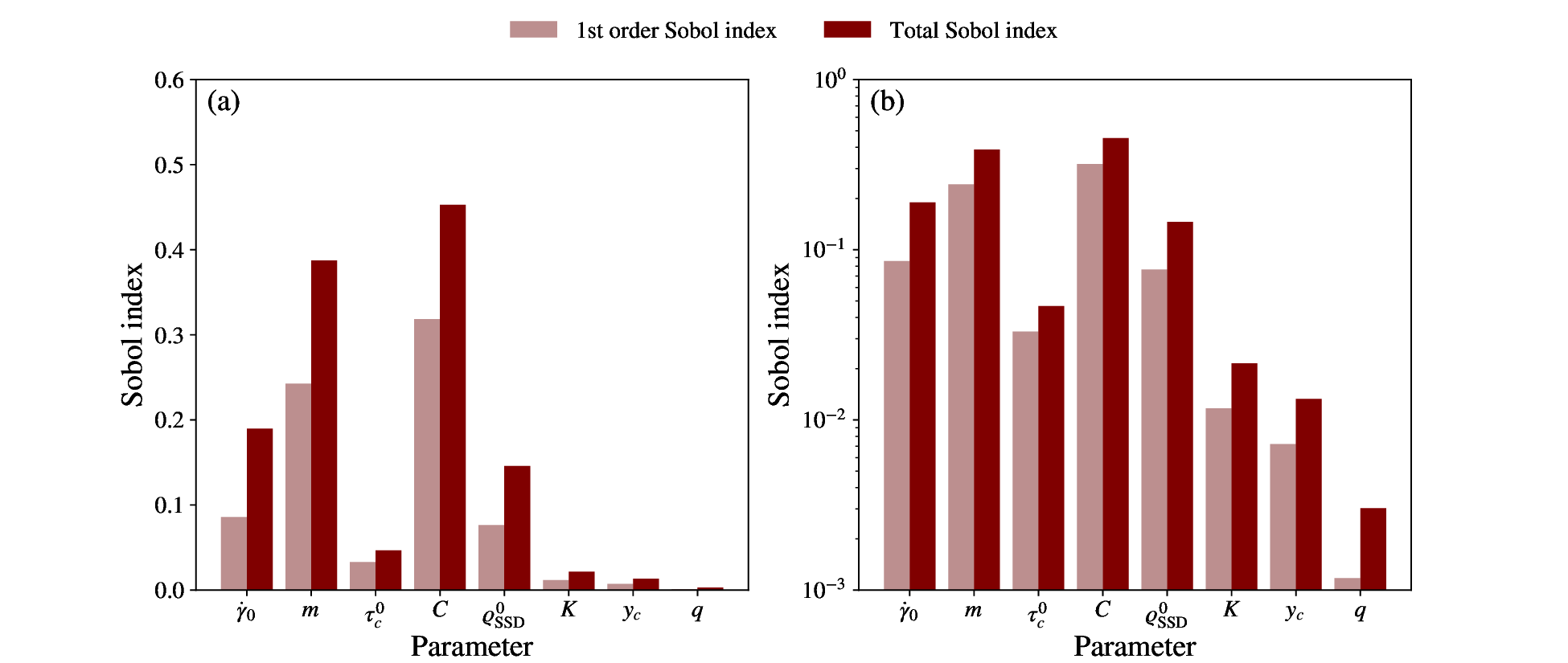}
    \caption{First order and total Sobol indices for the effect of crystal plasticity parameters on the difference between yield stress and maximum stress, $\sigma_{\mathrm{max}}-\sigma_y$, of the simulated tensile curve. Indices are shown on a) a linear scale and b) a logarithmic scale to show the difference between lower values.}
    \label{fig:Si_delta_stress}
\end{figure}

\begin{figure}[h!]
    \centering
    \includegraphics[width=1\textwidth]{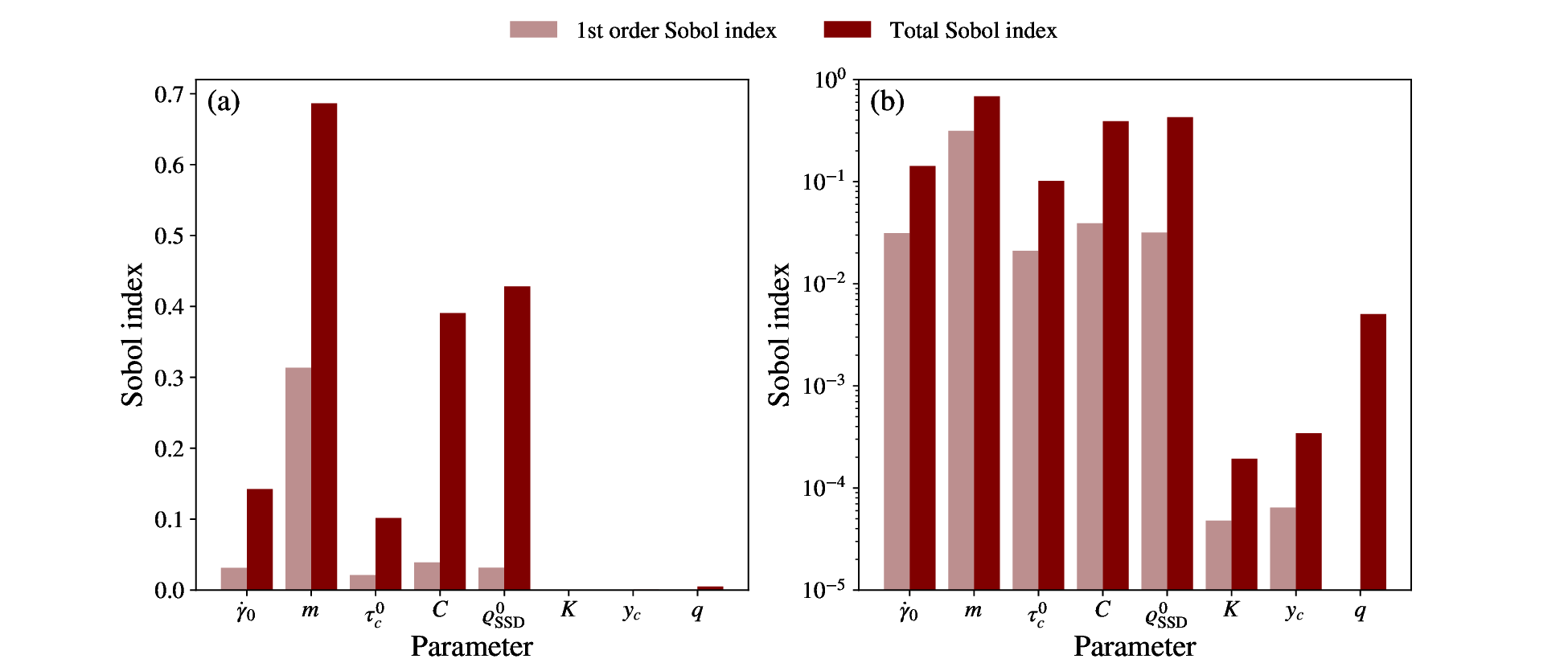}
    \caption{First order and total Sobol indices for the effect of crystal plasticity parameters on the normalised mean squared error with respect to experimental data. Indices are shown on a) a linear scale and b) a logarithmic scale to show the difference between lower values.}
    \label{fig:Si_res}
\end{figure}

The 1st order and total Sobol indices are shown for the influence of the crystal parameters on the yield stress, difference between yield stress and maximum stress, and the NMSE with respect to the experimental data in Figures \ref{fig:Si_yield_stress}, \ref{fig:Si_delta_stress} and \ref{fig:Si_res} respectively. 65~536 samples from the surrogate model were used for the analysis, as this gave minimal spread in the indices when the analysis was completed multiple times. This shows why using a surrogate model to compute the global Sobol sensitivity indices is crucial, as evaluating the crystal plasticity model this many times would be infeasible. The computational burden is instead reduced to the 320 crystal plasticity model evaluations needed to generate training data for the surrogate model.

From Equation \ref{eq:Flow_rule}, it can be expected that the rate insensitivity exponent, $m$, and critical resolved shear stress, $\tau_c$, will affect the point at which plastic slip will occur, and therefore the yield point of the stress strain curve. The critical resolved shear stress is related to the threshold shear stress, $\tau_c^0$, the geometric constant, $C$, and the initial SSD density, $\varrho_{\mathrm{SSD}}^0$, according to Equation \ref{eq:Hardening_rule}. This explains the sensitivity of the yield stress to $m$, $C$ and $\varrho_{\mathrm{SSD}}^0$ in Figure \ref{fig:Si_yield_stress}. 

In Figure \ref{fig:Si_yield_stress}, the total order sensitivity index of $C$ and $\varrho_{\mathrm{SSD}}^0$ are much larger than the 1st order index for each. While not shown here in entirety, the higher order indices also give further information about the sensitivity. For example, the combined effect of  $C$ and $\varrho_{\mathrm{SSD}}^0$ contributes between 16-18\% of the total sensitivity index for each suggesting a strong interdependence. This is expected as these parameters combine with a multiplicative effect to increase the critical resolved shear stress in Equation \ref{eq:Hardening_rule}. The yield stress is also much more sensitive to $C$ and $\varrho_{\mathrm{SSD}}^0$ than to  $\tau_c^0$, which is most likely attributed to the parameter bounds used in the sensitivity analysis. The total magnitude of the effect on critical resolved shear stress which can be provided by the initial dislocation density term is larger than that provided by the threshold shear stress meaning across their range, $C$ and $\varrho_{\mathrm{SSD}}^0$ will have a larger effect on the output. This demonstrates how the results of the sensitivity analysis are dependent on the parameter bounds used in the analysis. Additionally, parameters $K$ and $y_c$ controlling the evolution of SSD density are seen to have a near-zero effect on the yield stress. This is expected as the SSD density is expected to remain constant with a rate of zero until the onset of plasticity at the yield stress.

From Figure \ref{fig:Si_delta_stress} it can be seen that the difference between the yield stress and maximum stress is very dependent on the rate insensitivity exponent, $m$, and also on the geometric constant, $C$. The geometric constant multiplies the effect of dislocations in the hardening law in Equation \ref{eq:Hardening_rule} so is very influential on the rate of hardening in the plastic regime after yield. The effect of the mean free path hardening constant, $K$, and athermal annihilation radius, $y_c$ is now more influential in comparison to the results for their effect on the yield stress. These parameters control the evolution of SSDs in Equation \ref{eq:SSD_evolution} and therefore also have an influence on the hardening behaviour after yield. Similarly, Figure \ref{fig:Si_res} shows high sensitivity of the NMSE between simulation and experimental data to the rate insensitivity exponent, initial SSD density and the geometric constant. However, the sensitivity of NMSE on the crystal plasticity parameters can be expected to depend on which points on the stress strain curve are used to calculate the NMSE value. For example, if the simulation produces a higher density of output points in the elastic portion of the curve then the NMSE is weighted to simulations which correctly capture the elastic region and yield point. Another observation from Figure \ref{fig:Si_res}a is that the total Sobol index for every parameter is at least twice the magnitude of the first order index for each. This shows that there are considerable combined higher order effects between the model parameters and demonstrates the complex interactions between parameters in finding values which match closely to experimental data.

\subsection{Minimisation of the calibration objective function} \label{sec:results_calib}
Due to the sensitivity of the model outputs to the initial SSD density, $\varrho_{\mathrm{SSD}}^0$, and the rate insensitivity exponent, $m$, these parameters were selected for calibration in the optimisation procedure. The justification of including these parameters is that they affect the yield point of the curve and the behaviour of the curve in the transition between elastic and plastic behaviour. The mean free path hardening factor, $K$, is also chosen due to the effect on the hardening behaviour from controlling the rate of SSD accumulation. While this parameter had a lower effect overall on the quantities of interest than some of the other parameters, it is included because it is shown to only affect the tensile curve after yield. This is expected to help reduce the combined higher order interactions between the parameters being calibrated and simplify the calibration objective function somewhat. The values set for those parameters not included in the calibration process are shown in Table \ref{tab:calib}.

\begin{table}[h!]
\centering
\begin{tabular}{c p{0.25\textwidth} c p{0.40\textwidth}}
Parameter & Description & Magnitude & Remarks \\ \hline \hline

$\dot{\gamma_0}$ & \RaggedRight Initial slip rate ($s^{-1}$) & 0.001 & \RaggedRight Ref. \cite{gonzalez2014modelling}\\

$\tau_c^0$ & \RaggedRight Initial slip resistance (MPa) & 5.0 & \RaggedRight Set as a small value so hardening is primarily determined by the evolution of disloction densities.\\

$\alpha$ & \RaggedRight Geometric factor & 0.35 & Ref. \cite{ashby1970deformation} \\

$y_c$ & \RaggedRight Annihilation radius (mm) & $1.25\times 10^{-6}$ & Ref. \cite{plancher2019validity} \\

$q$ & \RaggedRight Latent hardening constant & 1.2 & \RaggedRight The ratio of self-hardening to latent hardening of non-coplanar systems is typically from 1.0-1.4 \cite{kocks1970relation}, so 1.2 is assumed here.\\

\end{tabular}

\caption{\label{tab:calib} Descriptions and values used for the fixed crystal plasticity parameters in the calibration procedures.}
\end{table}

\begin{figure}[h!]
    \centering
    \includegraphics[width=1\textwidth]{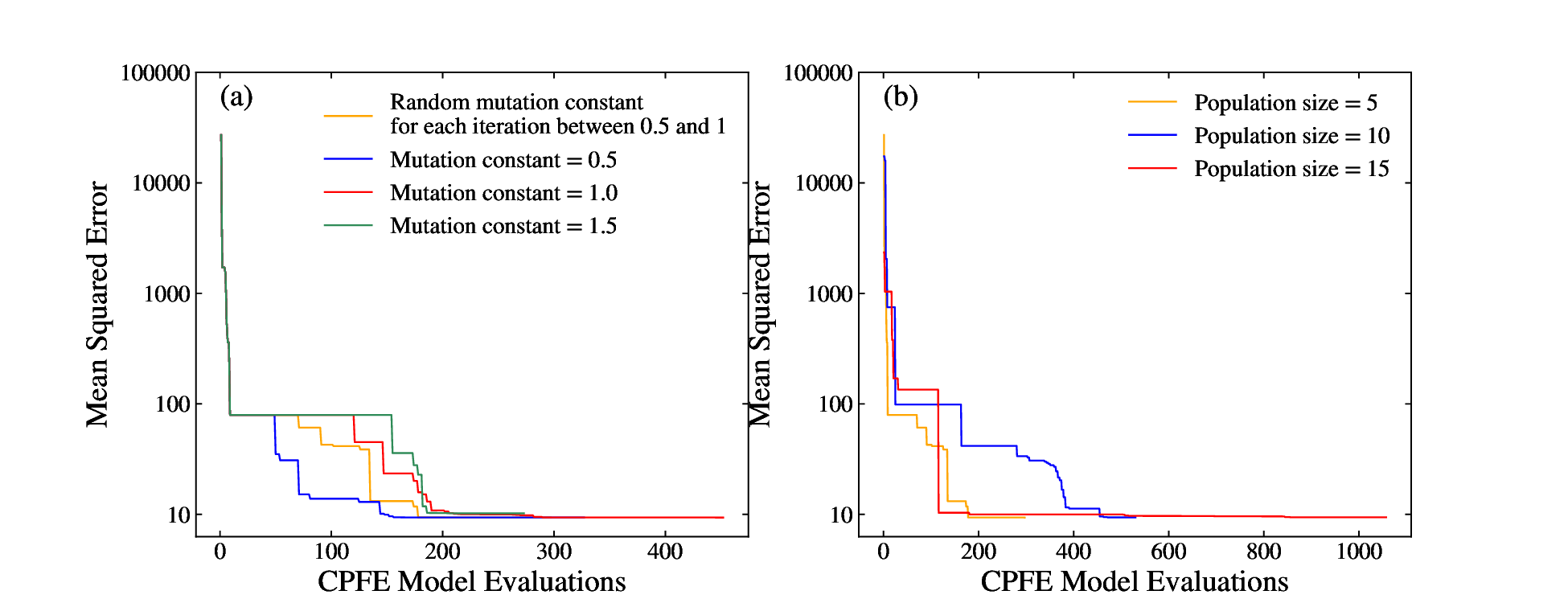}
    \caption{Convergence of the differential evolution optimisation algorithm with different optimisation hyperparameters. a) compares performance for different mutation constants and b) compares performance for different population sizes at each generation.}
    \label{fig:DE_conv}
\end{figure}

Figure \ref{fig:DE_conv} shows the convergence of the differential evolution algorithm for different optimisation hyperparameters. A random seed of 99 was set for all of the runs to allow a comparison that was not sensitive to the initial conditions. The final NMSE value and parameter values are summarised for each optimisation run in Table \ref{tab:DE_results}.

\begin{table}[h!]
    \centering
    \begin{tabular}{lcccc}
        Model Run & $m$ & $\varrho_{\mathrm{SSD}}^0$ & $K$ & Final NMSE \\ \hline \hline
        
        \begin{tabular}{@{}l@{}}Mutation constant = [0.5-1.0] \\ Population size = 5\end{tabular} &  0.470 & $3.31\times10^7$ & 0.0976 & 9.37\\

        \begin{tabular}{@{}l@{}}Mutation constant = 0.5 \\ Population size = 5\end{tabular} &  0.470 & $3.31\times10^7$ & 0.0975  & 9.37 \\

        \begin{tabular}{@{}l@{}}Mutation constant = 1.0 \\ Population size = 5\end{tabular} &  0.473 & $3.34\times10^7$ & 0.0973  & 9.36 \\

        \begin{tabular}{@{}l@{}}Mutation constant = 1.5 \\ Population size = 5\end{tabular} &  0.448 & $3.16\times10^7$ & 0.1004  & 10.22 \\

        \begin{tabular}{@{}l@{}}Mutation constant = [0.5-1.0] \\ Population size = 10\end{tabular} &  0.477 & $3.36\times10^7$ & 0.0975  & 9.40 \\

        \begin{tabular}{@{}l@{}}Mutation constant = [0.5-1.0] \\ Population size = 15\end{tabular} & 0.477 & $3.37\times10^7$ & 0.0719  & 9.41 \\

    \end{tabular}
    \caption{Final parameter sets for each run of the differential evolution algorithm with the final NMSE residual for each.}
    \label{tab:DE_results}
\end{table}

Figure \ref{fig:DE_conv}a shows the effect of different mutation constants on the optimisation convergence. All four runs were completed with a population size of 5 and a combination constant of 0.7. The lowest mutation constant of 0.5 converges most quickly to a global solution. This represents a search strategy which favours exploitation of the current estimate of the optimum over exploration further within the parameter space. While allowing the mutation constant to randomly dither between values in the range 0.5-1.0 gives better performance than setting a constant of 1.0, it is still outperformed by having a mutation constant of 0.5.

Similarly, Figure \ref{fig:DE_conv}b shows the effect of population size on optimisation performance. Again, a combination constant of 0.7 was used for the 3 runs and dithering was used to allow the mutation constant to take a random value between 0.5-1.0 for each iteration. The best performance is given by the smallest population size of 5, as this requires fewer model evaluations per iteration. However, the speed with which this finds the global optimum is only marginally faster than with a population size of 15 suggesting a non-monotonic relationship between efficiency and the population size. The effect of the combination constant was also investigated, but this was found to have very little impact on the search path of the algorithm in this instance.

\begin{figure}[h!]
    \centering
    \includegraphics[width=0.65\textwidth]{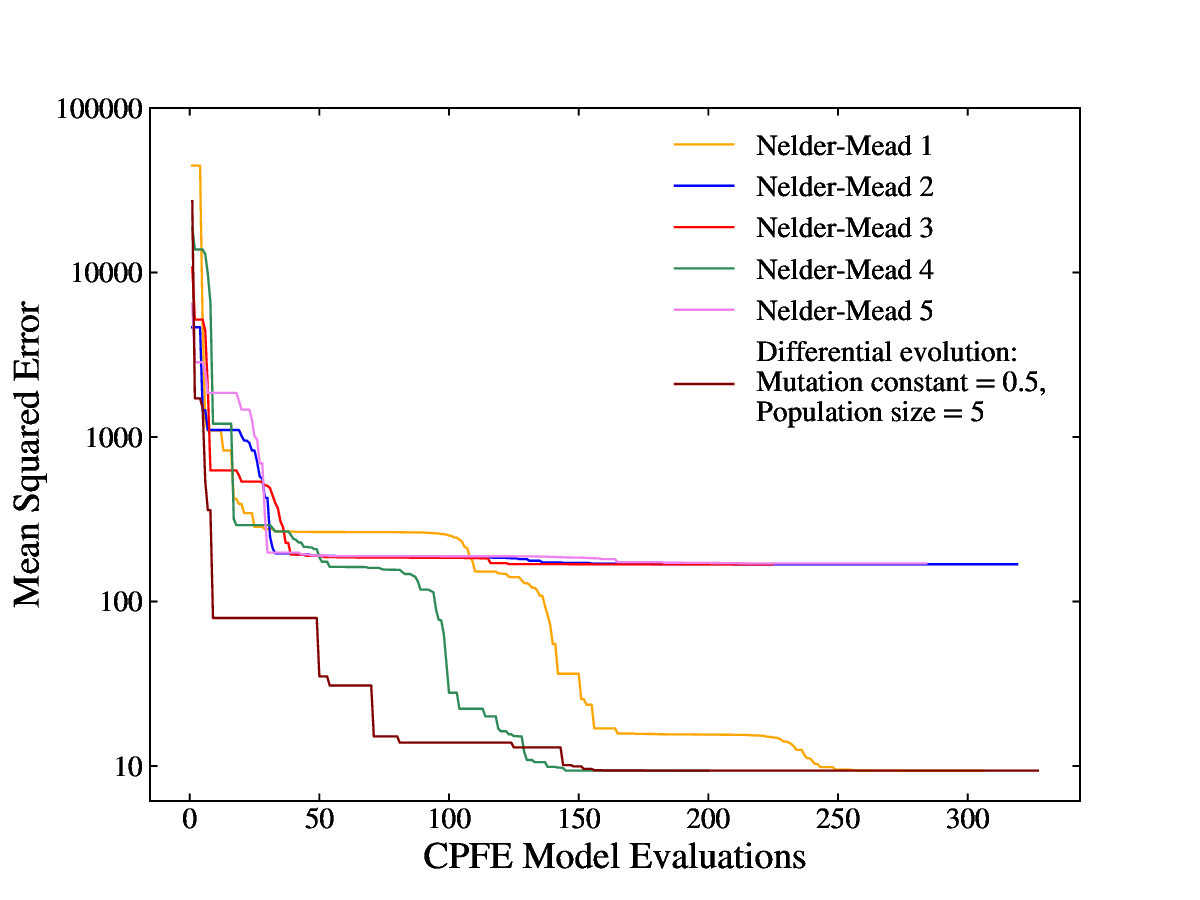}
    \caption{Convergence of the Nelder-Mead optimisation algorithm for 5 random starting locations. The best performing differential evolution run is shown for comparison.}
    \label{fig:NM_conv}
\end{figure}

The best performing differential evolution run with a population size of 5 and mutation constant of 0.5 is shown in Figure \ref{fig:NM_conv} alongside five runs of the Nelder-Mead algorithm. The final NMSE and final parameter set for each run are shown in Table \ref{tab:NM_results}.

\begin{table}[h!]
    \centering
    \begin{tabular}{lcccc}
        Model Run & $m$ & $\varrho_{\mathrm{SSD}}^0$ & $K$ & Final NMSE \\ \hline \hline
        
        \begin{tabular}{@{}l@{}l@{}}Differential evolution \\ Mutation constant = 0.5 \\ Population size = 5\end{tabular} &  0.470 & $3.31\times10^7$ & 0.0975  & 9.37 \\

        Nelder-Mead run 1 &  0.025 & $1.77\times10^7$ & 0.0741  & 168.14 \\

        Nelder-Mead run 2 &  0.026 & $1.77\times10^7$ & 0.0750  & 168.57 \\

        Nelder-Mead run 3 &  0.472 & $3.33\times10^7$ & 0.0974  & 9.36 \\

        Nelder-Mead run 4 &  0.472 & $3.33\times10^7$ & 0.0974  & 9.36 \\

        Nelder-Mead run 5 &  0.025 & $1.75\times10^7$ & 0.0772  & 170.58 \\

    \end{tabular}
    \caption{Final parameter sets for each run of the Nelder-Mead algorithm with the lowest NMSE residual shown for each. The best performing differential evolution run is shown for comparison.}
    \label{tab:NM_results}
\end{table}

Two of the Nelder-Mead runs converge to the same minimum as the differential evolution method which suggests the algorithm has been able to find the global minimum in these instances. However, the other three runs appear to get stuck in the same local minimum with an NMSE value around 170. The presence of local minima in the calibration shows that the objective function is non-convex and justifies the use of a global algorithm such as differential evolution. 

Despite one of the Nelder-Mead runs reaching the global minimum with fewer model evaluations than the differential evolution algorithm, multiple Nelder-Mead runs are clearly required and arriving at the global minimum is never assured with the local approach. Another crucial difference between the Nelder-Mead and differential evolution algorithm is the number of hyperparameters which can be tuned in the differential evolution algorithm, notably the population size and mutation constant. While these make the algorithm very adaptable to different problems, it also adds complexity in its usage as the most efficient hyperparameters can be expected to be problem specific and require user judgement to determine suitable values. This is reinforced in the difference in performance between various differential evolution runs shown in Figure \ref{fig:DE_conv}.

\begin{figure}[h!]
    \centering
    \includegraphics[width=0.65\textwidth]{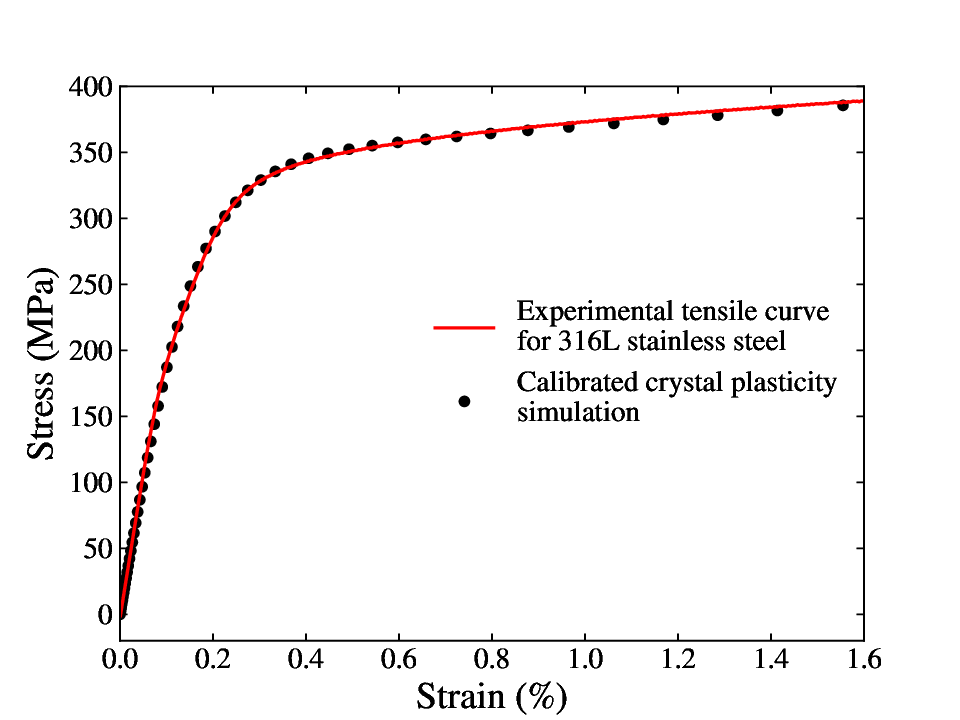}
    \caption{Comparison between the experimental stress strain curve and the simulated curve from the calibrated crystal plasticity model.}
    \label{fig:FinalCurve}
\end{figure}

Figure \ref{fig:FinalCurve} shows the experimental stress strain curve and the simulated curve corresponding to the parameter set found in the third and fourth run of the Nelder-Mead algorithm and the differential evolution algorithm. These parameters give a curve which shows very close agreement to the experimental data it has been calibrated to, which gives further confidence the optimisation has found the best set of parameters within the bounds provided. It also justifies the selection of these three parameters as the calibrated model gives close agreement with the experimental data.

\section{Conclusion} \label{sect:Conclusion}

Due to the number and complex interactions between parameters in crystal plasticity models, it is important to attempt to quantify and understand the effect of these parameters on output quantities of interest from the model. This study has demonstrated the utility of conducting a Sobol sensitivity analysis on the crystal plasticity model, not only in determining sensitivity of outputs to individual crystal plasticity parameters, but in determining the sensitivity to higher order combinations of parameters. However, this analysis required over 65~000 model evaluations so completing the sensitivity analysis on the crystal plasticity model itself would be prohibitively computationally expensive. For this reason, a Gaussian process regression model was required to emulate the response of the crystal plasticity model outputs with respect to the input parameters, allowing the sensitivity analysis to be conducted on this surrogate model.

The sensitivity analysis is able to inform on the most influential parameters which can be used to calibrate the crystal plasticity model with respect to experimental data for a specific material. A local Nelder-Mead optimisation and a global differential evolution algorithm have both been used to calibrate the model. Both the local method and global method have been shown to converge on a global optimum parameter set. However, the local method requires multiple restarts to find the global optimum due to the presence of local minima in the calibration objective function. Therefore, the global differential evolution algorithm is able to find the global optimum more reliably within a similar number of crystal plasticity model evaluations. The performance of the differential evolution algorithm was shown to be dependent on the optimisation hyperparameters selected by the user meaning care and judgement is required in selecting suitable values. The Nelder-Mead algorithm in contrast only requires a convergence criteria to be set by the user making its implementation simpler, while less efficient. Therefore, it is recommended that a global optimisation method, such as the differential evolution algorithm, is used for the calibration of crystal plasticity models to ensure consistent and repeatable identification of optimal crystal plasticity parameters. However, care should be taken in selecting the hyperparameters of such global algorithms to ensure good performance of the method.

A Gaussian process surrogate model has been shown to successfully predict outputs from the crystal plasticity model, particularly the NMSE residual. This provides a future avenue in utilising a Gaussian process surrogate model of the objective function to complete the calibration of crystal plasticity parameters in a Bayesian framework. The inherent quantification of model uncertainty provided by the Gaussian process could give an indication of the uncertainty in the parameter estimation as well as highlight areas where there is high uncertainty in the objective function prediction to give a more targeted search strategy.

\section*{Acknowledgements} \label{sect:Conclusion}
\noindent The authors gratefully acknowledge the support of UKRI through SINDRI (EP/V038079/1) as well as financial support from EDF Energy. Dr. Nicoló Grilli, University of Bristol, is gratefully thanked for the fruitful discussions. Prof. Mostafavi also acknowledges the support of Royal Academy of Engineering through a Research Chair Fellowship. Additionally, the authors acknowledge the support of the computational and data storage facilities of the Advanced Computing Research Centre, University of Bristol - http://www.bristol.ac.uk/acrc/.

\section*{Conflict of Interest Statement}
\noindent The authors declare that they have no known competing financial interests or personal relationships that could have appeared to influence the work reported in this paper.

\section*{Data Availability Statement}
\noindent The data that support the findings of this study are available from the corresponding author upon reasonable request.

\bibliographystyle{unsrt}
\bibliography{bibfile}

\end{document}